\documentclass[a4paper]{article}  
\usepackage[utf8]{inputenc}
\usepackage[T1]{fontenc}
\usepackage{lmodern}
\usepackage[scaled=.95]{libertine}
\usepackage{geometry}
\usepackage{gensymb}

\usepackage{graphicx}
\usepackage[protrusion=true, expansion=true, shrink=55, stretch=55, tracking=true, kerning=true, spacing=true, final]{microtype}
\usepackage{tabularx,multirow,booktabs}
\usepackage[super]{natbib}
\usepackage{relsize}             
\usepackage{tikz}
\usepackage{bm}
\usepackage{colortbl}
\usepackage{amsmath,amssymb} 
\usepackage{wrapfig}
\usepackage{scalefnt}
\usepackage[absolute,overlay]{textpos}
\definecolor{long}{HTML}{649522}
\definecolor{short}{HTML}{fe3e05}

\DeclareUnicodeCharacter{2212}{-}

\title{Insights gained from the Light Echo of Cepheus A HW2}
\author{Bringfried Stecklum \& Verena Wolf\\ Thüringer Landesternwarte Tautenburg}
\begin{document}
\maketitle
\section{Abstract}
Cepheus A HW2 is a well-studied massive young stellar object (MYSO) featuring Class II methanol masers$^1$. 
Recently, certain maser components have been found to flare every five years$^2$. 
This period went undetected in (NEO)WISE photometry due to detector saturation.
However, difference imaging
revealed a light echo (LE), representing a variability record. Its periodicity is similar to that of the masers. 
Thereby, the mid-IR light curve since 2007 could be reconstructed.
Phase and period maps provide information on the circumstellar dust distribution and viewing geometry. This is the first time an LE has been used for reverberation mapping of a YSO environment and expanding its light curve into the past.
\section{Data Acquisition and Analysis}
Mean W1 (3.4\,µm), and W2 (4.6\,µm) (NEO)WISE images for each of the 25 visits have been retrieved using ICORE$^3$. 
Subtracting the total average from the respective frames revealed the presence of a splendid LE,
shown in the below mosaic, where magenta indicates saturated pixels.
\begin{figure}[h]
    \begin{center}
        \includegraphics[width=0.925\textwidth]{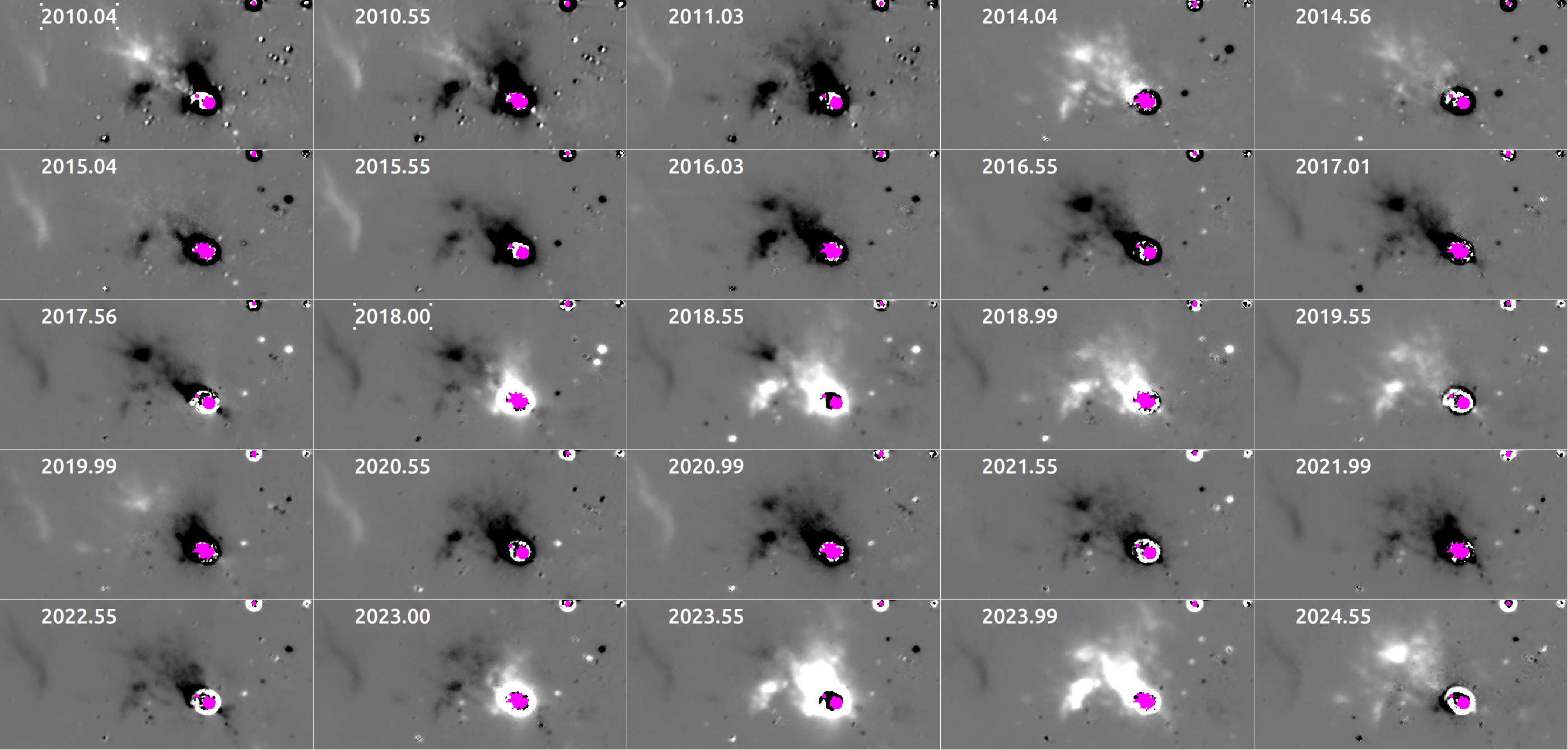}
    \end{center}
\end{figure}
Amplitude, phase, and period maps have been obtained from {\em pixel-wise} sine fits to the light curves for all pixels above the 1$\sigma$ noise level from the combined W1 and W2 data. 
\section{The complete (NEO)WISE Cep A HW2 Light Curve}
\begin{wrapfigure}[8]{r}{0.5\textwidth}
    \begin{center}
    \vspace*{-1cm}
        \includegraphics[width=.495\textwidth]{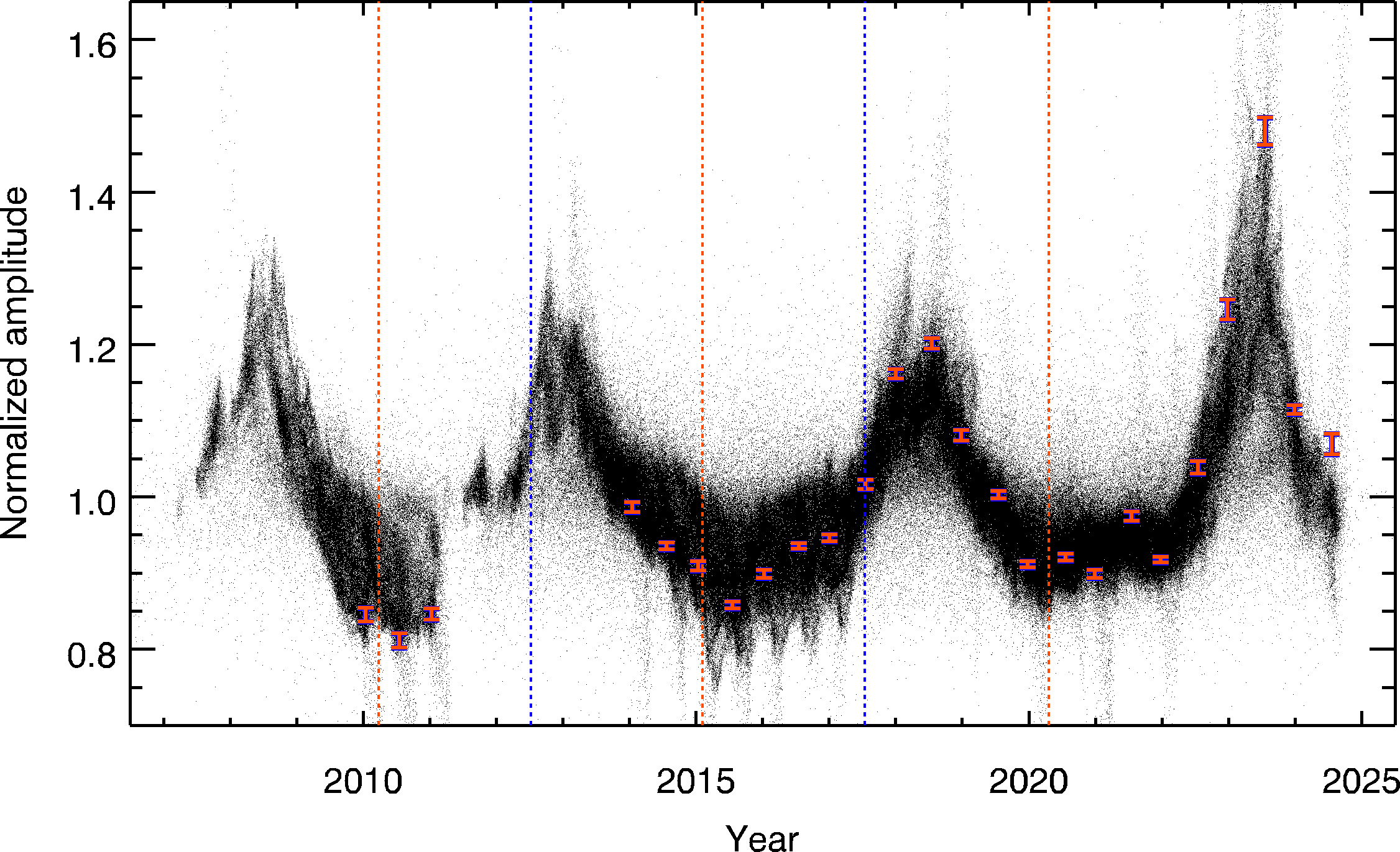}
    \end{center}
\end{wrapfigure}
The light curve, normalized to the mean flux, was restored from the LE by applying the individual phase shifts. It covers four peaks, separated by the 5\,yr period, while the (NEO)WISE photometry (red symbols) includes only the two most recent ones. The vertical red/blue dashed lines mark the peak dates of the red and blue shifted maser components, which show anticyclic behavior$^2$. 
\section{The Phase Map}
\begin{wrapfigure}[9]{r}{0.5\textwidth}
    \begin{center}
    \vspace*{-1cm}
        \includegraphics[width=.495\textwidth]{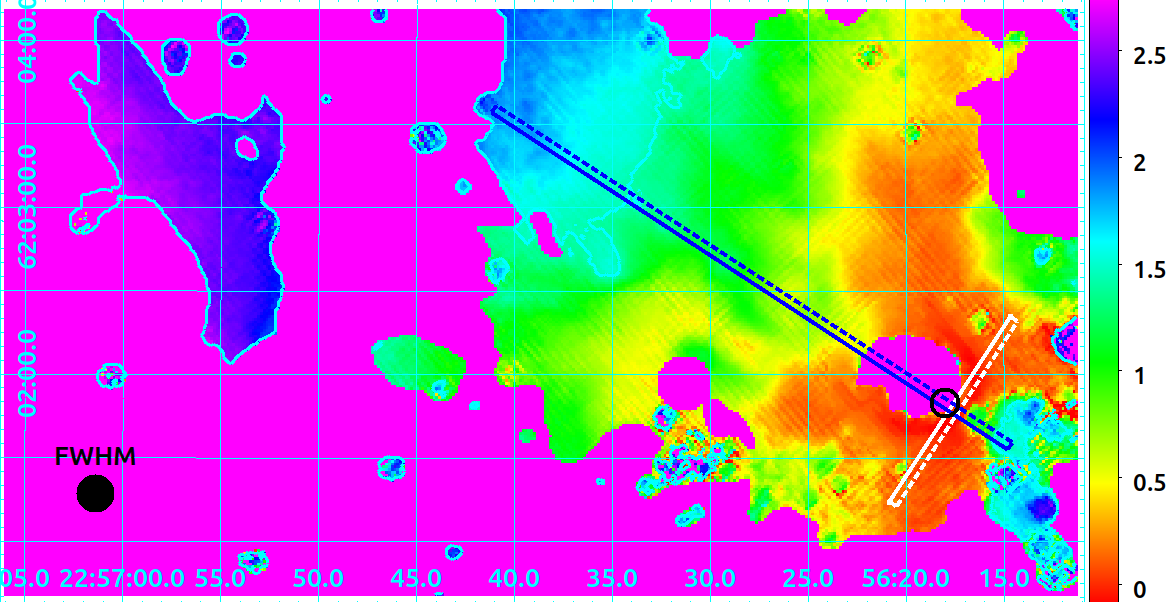}
    \end{center}
\end{wrapfigure}
The phase map shows the time delay [yr] of the LE. Magenta areas are below the noise threshold or saturated near Cep A HW2 (black circle). Most of the LE originates from the blueshifted north-eastern outflow lobe, whereas the redshifted part to the SW is more compact. The ratio of the phase gradients along the  NE and SW flow axis (blue) suggests a disk inclination of 51\degree$\pm$11\degree, consistent with previous estimates$^4$. 
The appearance of the phase distribution close to the MYSO is reminiscent of x-shaped outflow cavity scattering walls, which have a high column density. The phase profile along the disk major axis (white) is more complex with phase minima on either side of the source.

\section{The Period Map}
\begin{wrapfigure}[17]{r}{0.5\textwidth}
    \begin{center}
    \vspace*{-1cm}
        \includegraphics[width=.495\textwidth]{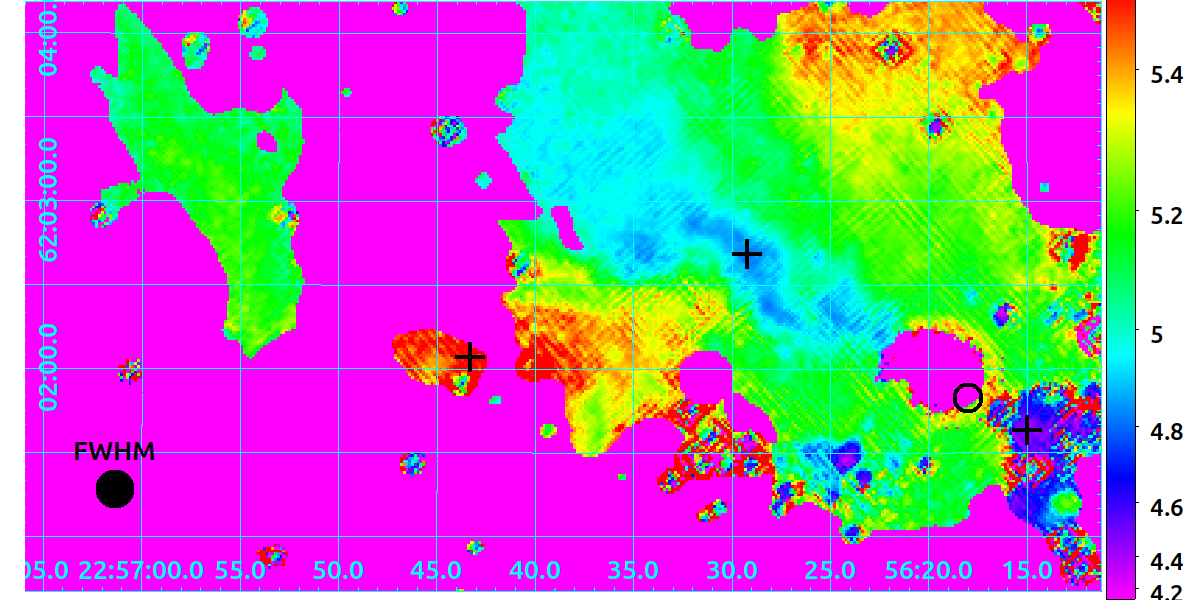}\\
    \vspace*{.25cm}
        \includegraphics[width=.48\textwidth]{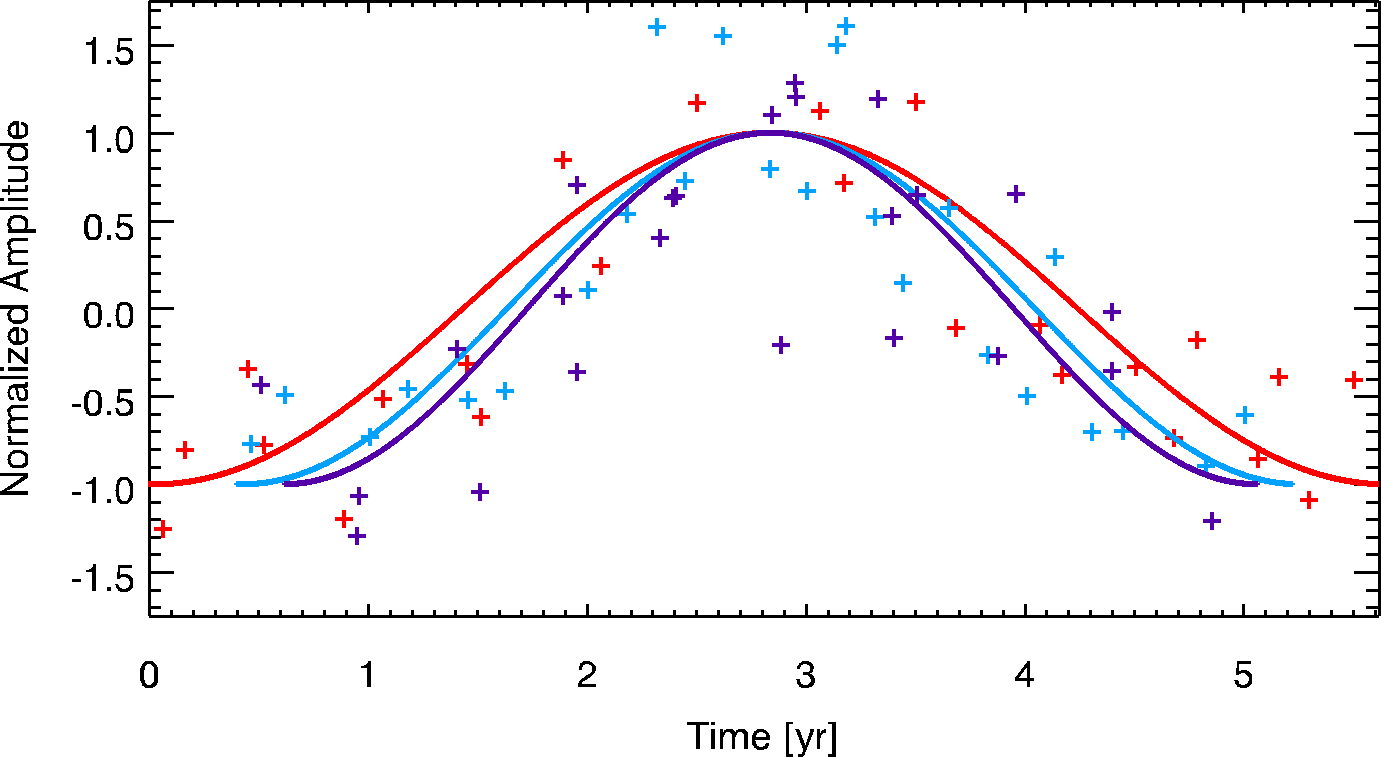}
    \end{center}
\end{wrapfigure}
Initially, the fixed maser period was applied in fitting the light curves. Adding the period to the free parameters gave a period map showing systematic variations. 
The shorter periods in the redshifted lobe and along the blueshifted outflow axis are
remarkable. Black crosses mark pixels for which the period-folded light curves are shown (lower panel) to illustrate the variations. They are yet unexplained and may arise from the interplay of geometric light-travel delays, multiple scattering-induced pulse broadening, and line-of-sight projection effects.

\section{Discussion}
Variability of the reflection nebula associated with Cep A HW2 in $K^\prime$ images had been found earlier$^5$.
It was attributed to orbiting dusty clumps at the outflow base, causing sweeping shadows across the nebula. This seems reasonable but does not explain the coherent spiky shape of the light curve. 
Protostellar pulsations$^6$ 
can be ruled out, as the 5 yr period would require a 100-fold luminosity of the MYSO.
The fact that this periodicity is only imprinted on low-flux masers$^2$ is puzzling. 
In the dense MYSO environment, single scattering does not hold. Instead, time-dependent multiple scattering simulations are required to model the LE, providing insights into the dust distribution around Cep A HW2. 
\vspace*{.5cm}
\\
{\bf\large References}\\
\small{[1] A. Sanna, et~al. (2025) {\em A\&A\/} 697:A206; [2] M. Durjasz, et~al. (2022) {\em A\&A\/} 663:A123; [3] F. Masci (2013) {\em ASCL\/} 02010; [4] N.~A. Patel, et~al. (2005) {\em Nature\/} 437:109; [5] K.~W. Hodapp, et~al. (2009) {\em AJ\/} 137:3501; [6] K. Inayoshi, et~al. (2013) {\em AJ\/} 769:L20.}
\end{document}